\documentclass[twocolumn, tighten, times]{aastex631}
\usepackage{graphicx,verbatim} 
\usepackage{soul}
\usepackage[T1]{fontenc}
\begin{document}
\title{H$_2^{18}$O in the terrestrial planet-forming regions of protoplanetary disks}

\correspondingauthor{Colette Salyk}
\email{cosalyk@vassar.edu}

\author[0000-0003-3682-6632]{Colette Salyk}
\affiliation{Vassar College, 124 Raymond Avenue, Poughkeepsie, NY 12604, USA}

\author[0000-0001-7552-1562]{Klaus M. Pontoppidan}
\affiliation{Jet Propulsion Laboratory, California Institute of Technology, 4800 Oak Grove Drive, Pasadena, CA 91109, USA}
\affiliation{Division of Geological and Planetary Sciences, California Institute of Technology, MC 150-21, Pasadena, CA 91125, USA}

\author[0000-0002-0661-7517]{Ke Zhang}
\affiliation{Department of Astronomy, University of Wisconsin-Madison, Madison, WI 53706, USA}

\author{Sophie Heinzen}
\affiliation{Vassar College, 124 Raymond Avenue, Poughkeepsie, NY 12605, USA}

\author[0000-0002-0150-0125]{Jenny K. Calahan}
\affiliation{Center for Astrophysics, Harvard \& Smithsonian, 60 Garden St., Cambridge, MA 02138, USA}

\author[0000-0003-4335-0900]{Andrea Banzatti}
\affiliation{Department of Physics, Texas State University, 749 N Comanche Street, San Marcos, TX 78666, USA}

\author[0000-0002-4555-5144]{D. Annie Dickson-Vandervelde}
\affiliation{Weber State University, Department of Physics,
1415 Edvalson St., Dept. 2508, Ogden, UT 84408-2508, USA}

\author[0000-0003-4179-6394]{Edwin A. Bergin}
\affiliation{Department of Astronomy, University of Michigan, 1085 S. University, Ann Arbor, MI 48109, USA}

\author[0000-0003-0787-1610]{Geoffrey A. Blake}
\affiliation{Division of Geological and Planetary Sciences, California Institute of Technology, MC 150-21, Pasadena, CA 91125, USA}

\author[0000-0003-2631-5265]{Nicole Arulanantham}
\affiliation{Astrophysics \& Space Institute, Schmidt Sciences, New York, NY 10011, USA}

\author[0000-0002-3291-6887]{Sebastiaan Krijt}
\affil{Department of Physics and Astronomy, University of Exeter, Exeter, EX4 4QL, UK}

\author[0000-0002-6695-3977]{John Carr}
\affiliation{Department of Astronomy, University of Maryland, 4296 Stadium Dr., College Park, MD 20742, USA}

\author[0000-0002-5758-150X]{Joan Najita}
\affiliation{NSF’s NOIRLab, 950 N. Cherry Avenue, Tucson, AZ 85719, USA}

\author[0000-0003-1665-5709]{Joel Green}
\affiliation{Space Telescope Science Institute, 3700 San Martin Drive, Baltimore, MD 21218, USA}

\author[0000-0001-7152-9794]{Carlos Romero-Mirza}
\affiliation{Center for Astrophysics, Harvard \& Smithsonian, 60 Garden St., Cambridge, MA 02138, USA}

\begin{abstract}
Isotopologues play an important role in solar system cosmochemistry studies, revealing details of early planet formation physics and chemistry.  Oxygen isotopes, as measured in solar system materials, reveal evidence for both mass-dependent fractionation processes and a mass-independent process commonly attributed to isotope-selective photodissociation of CO in the solar nebula. The sensitivity of JWST's MIRI-MRS enables studies of isotopologues in the terrestrial planet-forming regions around nearby young stars.  We report here on a search for H$_2^{18}$O in 22 disks from the JDISC Survey with evidence for substantial water vapor reservoirs, with the goal of measuring H$_2^{16}$O/H$_2^{18}$O ratios, and potentially revealing the predicted enhancement of H$_2^{18}$O caused by isotope-selective photodissociation. We find marginal detections of H$_2^{18}$O in six disks, and a more significant detection of H$_2^{18}$O in the disk around WSB 52.  Modeling of the detected H$_2^{18}$O lines assuming an ISM ratio of H$_2^{16}$O/H$_2^{18}$O predicts H$_2^{18}$O features consistent with observations for four of the modeled disks, but stronger H$_2^{18}$O features than are observed in three of the modeled disks, which includes WSB 52. Therefore, these latter three disks require a higher H$_2^{16}$O/H$_2^{18}$O ratio than the ISM in the water-emitting region, in contrast to long-standing theoretical expectations.  We suggest that either the H$_2^{18}$O-rich water has been removed from the emitting region and replaced by H$_2^{18}$O-poor water formed by reactions with $^{18}$O-poor CO, or that the gas-phase water is depleted in $^{18}$O via mass-dependent fractionation processes at the water snowline.
\end{abstract}

\section{Introduction}
Planets form in protoplanetary disks, and inherit the chemical conditions of their local disk environments.  The standard ``solar nebula'' hypothesis of planet formation predicts chemistry that comes to equilibrium according to the local temperature \citep{Grossman72}. However, both solar system data and a growing understanding of protoplanetary disks suggests that planet-formation chemistry may not be fully predictable; it may include contributions from materials that have not come to local equilibrium \citep[e.g.][]{Lodders05, Visser11}, and may be blended by radial transport \citep[e.g.][]{Cyr98,Brownlee06}.

The chemistry of water is particularly important to understand in the context of planet formation.  Water is a primary molecular carrier of oxygen \citep[e.g.][]{Pontoppidan14}, and when it condenses at the so-called ``snowline'', solid mass may approximately triple, facilitating the growth of larger cores and gas giant planets \citep{Hayashi81,Pollack96}.  Beyond the snowline, water ice is likely the dominant component of mm--cm-sized pebbles, which readily drift in the radial direction due to disk pressure gradients \citep{Pinilla12}.  For a uniform disk, the pressure gradient moves pebbles radially inwards \citep{Weidenschilling77}, and the migration and subsequent sublimation of the grains can enrich the inner disk in oxygen \citep[e.g.][]{Ciesla06,Booth17,Kalyaan21,Houge25}.  

In solar system cosmochemistry, {\it isotopologues} of water are used as a tracer of late stage migration, as isotopologue ratios are suggested to be correlated with radial origin \citep{Clayton73,McKeegan11,Genda16}.  Studies of water isotopologues have thus been used to trace the origins of Earth's oceans \citep[e.g.][]{Sarafian14,Bockelee-Morvan15,Wu18}.  Oxygen isotopologue ratios, in particular, require the action of mass-independent fractionation {\it in addition to} mass-dependent processes, with a leading hypothesis being that the mass-independent fractionation arises from isotope selective photodissociation \citep{Thiemens83,Yurimoto04,Lyons05}.  In short,  C$^{16}$O readily self-shields, while heavy isotopologues of CO photo-dissociate, enriching free oxygen with heavy isotopes.  These heavy oxygen isotopes then combine with hydrogen to form isotopically-heavy water. 

Studies of several molecules (including HCN, CN,  C$_2$H and CO) in the outer regions of protoplanetary disks support the selective photodissociation theory \citep{Smith09,Hily-Blant17,Hily-Blant19,Yoshida22,Bergin24}.  However, it's not yet known whether water in protoplanetary disks is affected by this mass-independent fractionation process, and, therefore, whether water isotopologues may serve as origins tracers in exoplanetary systems.  \citet{Calahan22} modeled this selective photodissociation process and asked whether the effects on water isotopologues might be detectable with infrared spectroscopy.  Indeed, they found an enhancement in H$_2^{18}$O above the ISM (i.e., a lower H$_2^{16}$O/H$_2^{18}$O ratio) by a factor of $\sim$2 in the infrared line-emitting region of their model.

The sensitivity and spectral resolution provided by the James Webb Space Telescope Mid InfraRed Instrument's Medium Resolution Spectrometer (JWST MIRI-MRS; \citealp{Rieke15,Wells15}) is greatly improving our ability to detect trace molecular species \citep[e.g.][]{Perotti23} and isotopologues \citep[e.g.][]{Grant23,Salyk25} in protoplanetary disks compared with past facilities (e.g., Spitzer-IRS; \citealp{Houck04}).  In this work, we leverage the large sample of disk spectra observed in Cycle 1 by the JWST Disk Infrared Spectral Chemistry Survey (JDISCS) team to search for H$_2^{18}$O.  We report several tentative detections, and one strong detection, of H$_2^{18}$O, in these spectra.  We model the properties of the H$_2^{18}$O, including its relative abundance, and discuss the implications for disk chemistry.

\section{Observations and Data Reduction}
Our investigation focuses on a subset of MIRI-MRS \citep{Wells15} spectra of Cycle 1 targets from the JWST Disk Infrared Spectral Chemistry Survey (JDISCS; \citealp{Pontoppidan24,Arulanantham25}). All of the raw data used in this paper can be found in MAST: \dataset[10.17909/hx6h-qw97]{http://dx.doi.org/10.17909/hx6h-qw97}. In order to detect isotopolgues of water, water optical depths must be high; we therefore limit ourselves to the subset of disks for which the water line/continuum ratio defined in \citet{Arulanantham25} is $\geq$0.05.  The list of targets is provided in Table \ref{table:disks}.
A full description of the observations (visit ID's, exposure times, observation dates/times) for these spectra can be found in \citet{Arulanantham25}.  Spectra presented in this work come from Cycle 1 programs 1584 (PI's: Salyk, Pontoppidan), 1640 (PI: Banzatti) and 1549 (PI: Pontoppidan).  The targets in program 1549, in particular, were selected due to their strong water emission as observed by Spitzer-IRS \citep{Pontoppidan10b}. They were observed with exposure times intended to push the limits of signal-to-noise ratios (SNRs) on water-rich spectra, with the intent of detecting trace molecules, including isotopologues of water.

JDISCS utilizes the JWST pipeline reduction (in this case, version 1.18.0 with CRDS 12.1.5; \citealp{Bushouse25}) up to stage 2b, and then utilizes customized routines to background subtract and fringe-correct using both stellar and asteroid fringe calibrators.  A detailed description of this process is provided in \citet{Pontoppidan24} and \citet{Arulanantham25}.   An additional improvement has since been added to the standard JDISCS calibration process: instead of treating the asteroids as blackbodies, they are modeled with a parametric emissivity (including silicate emission) adapted from \citet{Humes24}.  Model parameters are adjusted to minimize differences between standard stars HR 6538 and Mu Col reduced with this method, and their CALSPEC \citep{Bohlin20,Bohlin22} model spectra.  The JDISCS process can provide SNR improvements of a factor of $\sim$6 in channel 4 \citep{Pontoppidan24} which, as we will show, covers the wavelengths of interest for detection of H$_2^{18}$O. While the fringe-removal process appears to recover predicted statistical SNRs \citep{Pontoppidan24}, it is likely that residuals from fringe correction remain a substantial contributor to the overall noise budget.  Thus, we give careful consideration to fringe-correction residuals in the subsequent analysis in this work.

To analyze molecular emission, all spectra are continuum-subtracted as described in \citet{Banzatti25}.  In short, we utilize an iterative process in which a smoothed spectrum is created, and points below the smoothed spectrum are used to make a new smoothed spectrum \citep{Pontoppidan24}.  In a final step, regions expected to be line free are used to apply a wavelength-dependent flux offset.  This empirical continuum determination approach can sometimes cause blended line emission to be interpreted as continuum, so we interpolate across the 13.4--14.1 $\mu$m organic emitting region; however, this region will not play a role in the analysis presented here.  As we will show, the dominant isotopologue features of interest lie in the 22--28 $\mu$m range, where the continuum is readily distinguished from narrow gas-phase emission lines.

\section{Analysis}
\subsection{H\textsubscript{2}\textsuperscript{16}O models}
\label{sec:h216o_models}
We begin our analysis with modeling of the H$_2^{16}$O emission.  The reasoning for this is two-fold.  Firstly, H$_2^{16}$O emission blankets the MRS spectra throughout the full 5--28 $\mu$m spectral range, so its properties must be well-understood to detect or analyze the much weaker emission from isotopologues.  Secondly, understanding the properties of the water reservoir allows us to predict the strongest H$_2^{18}$O emission lines.

We determine the H$_2^{16}$O emission properties by fitting continuum-subtracted spectra with slab models. The spectral range from 13--27\,$\mu$m is modeled simultaneously.  H$_2^{16}$O is represented by the sum of three temperature components --- hot, warm, and cold --- defined by priors on temperature between 500–1500\,K, 200–800\,K, and 100–400\,K, in that order. Each component has three free parameters: excitation temperature, $T$, column density, $N$, and emitting area, $A$.  While three-component models are not as physically realistic as models that, for example, incorporate radial gradients, even two-temperature models capture much of the water emission details \citep{Romero-Mirza24}, and three-component models appear to reproduce spectra as well as more complex models \citep{Temmink24}. We further discuss the potential implications of this modeling choice in Section \ref{sec:anomalous}.  The cold H$_2$O component is limited to the 18---27\,$\mu$m range, as it contributes negligibly to shorter wavelengths.  OH is modeled with a single temperature component and the same set of free parameters. Synthetic spectra are generated using the \texttt{spectools\_ir} package \citep{Salyk22}\footnote{\url{https://github.com/csalyk/spectools_ir}}, using molecular data from HITRAN \citep{Gordon22}.  We assume a Gaussian line profile, with instrumental FWHM values of 110\,km/s for 13---18\,$\mu$m and 130\,km/s for 18---27\,$\mu$m, consistent with the resolving power measured by \citet{Banzatti25}. In addition to instrumental broadening, we include thermal broadening based on the excitation temperature of each molecular component.  Flux uncertainties are estimated using the Gaussian process procedure more fully described in \citet{Romero-Mirza24}.  In short, the noise is described by the sum of a slowly-varying component with length scale equal to two times the resolution, and a white-noise component based on pixel-to-pixel variations.  The slowly-varying component is assumed to be a Gaussian of amplitude 2 mJy for 13--18 $\mu$m, and 4 mJy for 18--27$\mu$m, and the magnitude of the white-noise component is determined by minimizing the negative log-likelihood.

Parameter estimation is performed using the Markov Chain Monte Carlo (MCMC) ensemble sampler \texttt{emcee} \citep{Foreman-Mackey13}, typically with 70 walkers and 3000–5000 steps. The chains generally converge within the first 2000 steps, and the remaining samples are used to compute posterior distributions. The best-fitting parameters for the slab models are provided in Table \ref{table:disks}.  Further details will be provided in Zhang et al., 2025, in preparation.

Before investigating water isotopologue emission, we produce new model spectra using a more extended linelist derived from the HITEMP database \citep{Rothman10}, rather than the HITRAN database, which, for user convenience, is restricted to more strongly allowed and lower-excitation level lines. (The HITEMP database was not used in the water fitting process itself, as the size of the database significantly slows model computations, making exploration of a large parameter space too time intensive.) We find this step necessary as some high-temperature lines excluded from HITRAN can be as strong as expected isotopologue emission lines --- see Appendix \ref{sec:hitemp_hitran_comparison}.  For our intermediate-size linelist, we extract lines from HITEMP with upper level energy $<=$15,000 K and Einstein A-coefficient $>10^{-3}$ s$^{-1}$.  Select portions of the MRS continuum-subtracted spectra and 3-component models produced using the HITEMP database  are shown in Figures \ref{fig:allspectra_short} and \ref{fig:allspectra_long} (for sources with marginal H$_2^{18}$O detections --- see Section \ref{sec:detections}) or in Figures \ref{fig:nondetections_short} and \ref{fig:nondetections_long} (for sources with non-detections of  H$_2^{18}$O).  Figure \ref{fig:allspectra_long_zoom} shows a portion of Figure \ref{fig:allspectra_long} zoomed in on the H$_2^{18}$O-emitting region.

\subsection{Potential detections of H\textsubscript{2}\textsuperscript{18}O}
\label{sec:detections}
Utilizing the best-fit H$_2^{16}$O models, we produce H$_2^{18}$O models assuming that both isotopologues arise from the same reservoirs, and that the 
H$_2^{16}$O/H$_2^{18}$O ratio is equal to the ISM $^{16}$O/$^{18}$O ratio of 557 \citep{Milam05}.  We discuss the implications of the same-reservoir assumption, and perform fits to the H$_2^{16}$O/H$_2^{18}$O ratio in later analysis sections.

\begin{figure*}[h]
\centering
\includegraphics[width=160mm]{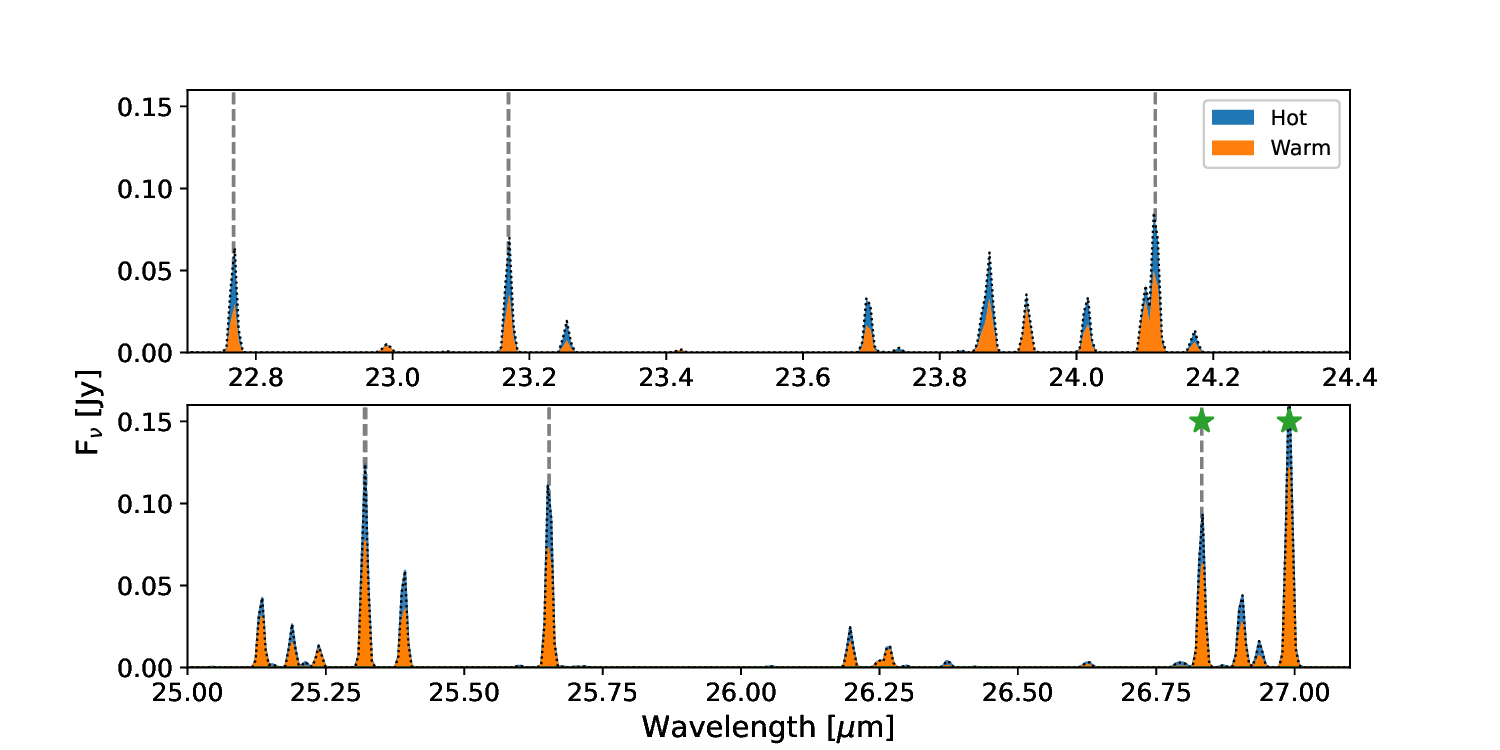}
\caption{Portions of the nominal H$_2^{18}$O emission model for WSB 52, assuming the emission arises in the same 3-component reservoirs as given in Table \ref{table:disks}, and that H$_2^{16}$O/ H$_2^{18}$O=557.  Blue and orange shading shows the contributions from the hot and warm reservoirs, respectively.  The cold contribution is not sufficient to be visible on this scale.  The dotted curve shows the sum of the three components.  Dashed vertical lines indicate the emission lines selected for Table \ref{table:stronglines}; green stars mark the two features used while fitting the H$_2^{16}$O/H$_2^{18}$O ratio.
\label{fig:contribution_plot}}
\end{figure*}

Figure \ref{fig:contribution_plot} shows example model H$_2^{18}$O emission spectra for one target, WSB 52.  The strongest H$_2^{18}$O emission features lie in the 22--28 $\mu$m region. For WSB 52, and for nearly all disks in our sample, the warm component dominates the H$_2^{18}$O emission features in the model, and the cold component makes a negligible contribution.  For DoAr 25 and HT Lup, the hot component dominates slightly over the warm component.  This suggests that any detected H$_2^{18}$O emission is likely to reflect the inner hot and warm gas, rather than a colder component that has been proposed to be enriched by pebble drift \citep{Banzatti23b}. 

Table  \ref{table:stronglines} lists the H$_2^{18}$O features found to be strongest in the modeled spectra.  We find significant, though not uniform, overlap with the strongest lines found via thermochemical modeling by \citet{Calahan22}.  It is important to note that most of the strong transitions we identify come in closely-spaced ortho (parallel spin state)/para (opposite spin state) pairs, with corresponding ortho/para upper level degeneracy ratios of 3. So, in the optically thin regime, the weaker para transition should not be ignored, as it adds additional flux at the $\sim$30\% level.

\begin{rotatetable*}
\begin{deluxetable*}{l c c c c c c c c c c c c c c}
\tablecaption{List of JDISCS targets with detected water vapor.\label{table:disks}}
 \tabletypesize{\small}
\tablehead{
\colhead{Disk} & \colhead{SNR \tablenotemark{a} }&\colhead{P/C \tablenotemark{b}} &\colhead{T$_\mathrm{hot}$\tablenotemark{c}} &\colhead{log N$_\mathrm{hot}$} & \colhead{log A$_\mathrm{hot}$} &\colhead{T$_\mathrm{warm}$} &\colhead{log N$_\mathrm{warm}$} & \colhead{log A$_\mathrm{warm}$} &\colhead{T$_\mathrm{cold}$} &\colhead{log N$_\mathrm{cold}$} & \colhead{log A$_\mathrm{cold}$} &\colhead{T$_\mathrm{OH}$} &\colhead{log N$_\mathrm{OH}$} & \colhead{log A$_\mathrm{OH}$} \\
&&&\colhead{(K)}&\colhead{(cm$^{-2}$)}&\colhead{(AU$^{2}$)}&\colhead{(K)}&\colhead{(cm$^{-2}$)}&\colhead{(AU$^{2}$)}&\colhead{(K)}&\colhead{(cm$^{-2}$)}&\colhead{(AU$^{2}$)}&\colhead{(K)}&\colhead{(cm$^{-2}$)}&\colhead{(AU$^{2}$)} }
\startdata
AS 205 N  &  3.86  &  0.22  &  886  &  18.3  &  0.145  &  420  &  18.3  &  1.344  &  175  &  16.7  &  2.940  &  1826  &  14.4  &  1.432 \\
CI Tau  &  0.70  &  0.27  &  1039  &  18.0  &  -0.605  &  604  &  17.9  &  -0.016  &  297  &  16.9  &  0.798  &  1913  &  13.5  &  1.737 \\
DoAr 25  &  4.70  &  0.14  &  692  &  18.3  &  -0.867  &  473  &  18.0  &  -0.294  &  195  &  14.9  &  2.739  &  1275  &  13.8  &  0.965 \\
DoAr 33  &  2.01  &  0.11  &  727  &  18.2  &  -1.014  &  419  &  17.8  &  -0.199  &  225  &  15.4  &  1.380  &  1198  &  13.7  &  1.300 \\
Elias 2-20  &  1.09  &  0.34  &  828  &  18.1  &  -0.233  &  426  &  17.7  &  0.976  &  207  &  15.7  &  2.933  &  1869  &  13.6  &  1.846 \\
Elias 2-24  &  4.15  &  0.29  &  792  &  18.2  &  0.135  &  449  &  17.8  &  1.237  &  228  &  16.6  &  2.440  &  1707  &  13.9  &  1.695 \\
Elias 2-27  &  4.55  &  0.54  &  841  &  18.2  &  -0.561  &  464  &  18.1  &  0.361  &  201  &  15.9  &  2.448  &  1988  &  13.7  &  1.289 \\
FZ Tau  &  4.48  &  0.53  &  892  &  18.3  &  -0.161  &  472  &  18.2  &  0.760  &  226  &  16.3  &  1.958  &  2012  &  13.6  &  1.708 \\
GK Tau  &  1.92  &  0.10  &  947  &  17.9  &  -0.751  &  466  &  17.6  &  0.397  &  187  &  15.9  &  2.824  &  1778  &  13.5  &  1.847 \\
GO Tau  &  3.65  &  0.06  &  712  &  18.0  &  -1.600  &  489  &  17.3  &  -0.757  &  233  &  15.1  &  1.390  &  1301  &  14.5  &  0.006 \\
GQ Lup  &  1.27  &  0.12  &  895  &  17.7  &  -0.561  &  433  &  17.4  &  0.840  &  215  &  16.1  &  2.444  &  1529  &  13.9  &  1.723 \\
HP Tau  &  2.16  &  0.05  &  795  &  18.0  &  -0.454  &  531  &  17.2  &  0.303  &  170  &  16.2  &  2.752  &  1601  &  13.8  &  1.758 \\
HT Lup A+B  &  0.00  &  0.06  &  823  &  18.3  &  -0.438  &  465  &  18.0  &  0.039  &  179  &  17.5  &  1.065  &  1904  &  14.1  &  1.039 \\
IQ Tau  &  1.41  &  0.15  &  1009  &  18.1  &  -1.022  &  565  &  17.7  &  -0.379  &  193  &  15.6  &  2.752  &  1980  &  13.9  &  1.071 \\
IRAS 04385+2550  &  1.51  &  0.15  &  591  &  18.4  &  -0.273  &  358  &  18.1  &  0.704  &  174  &  16.2  &  2.942  &  1510  &  14.2  &  0.692 \\
RU Lup  &  3.57  &  0.19  &  1023  &  18.2  &  -0.257  &  572  &  18.0  &  0.705  &  231  &  16.0  &  2.483  &  1743  &  13.8  &  1.884 \\
SR 4  &  0.00  &  0.10  &  1006  &  17.8  &  -0.769  &  695  &  17.5  &  -0.222  &  198  &  16.9  &  1.185  &  1770  &  13.7  &  1.653 \\
Sz 114  &  3.24  &  0.26  &  695  &  18.4  &  -0.594  &  418  &  17.9  &  0.276  &  197  &  15.5  &  2.956  &  1282  &  14.7  &  0.440 \\
Sz 129  &  2.17  &  0.24  &  941  &  17.8  &  -0.949  &  521  &  17.4  &  0.092  &  190  &  15.5  &  2.889  &  1720  &  13.5  &  1.628 \\
TW Cha  &  4.25  &  0.47  &  905  &  17.9  &  -0.424  &  508  &  17.5  &  0.544  &  201  &  15.7  &  2.948  &  1658  &  13.6  &  1.767 \\
VZ Cha  &  2.78  &  0.38  &  1033  &  17.8  &  -0.455  &  609  &  17.5  &  0.373  &  227  &  16.1  &  2.289  &  1806  &  13.8  &  1.743 \\
WSB 52  &  11.06  &  0.73  &  659  &  18.6  &  0.235  &  400  &  18.6  &  1.104  &  167  &  16.4  &  2.832  &  1568  &  13.6  &  1.611 \\
\enddata

\tablenotetext{a}{Calculated SNR ratio of the 26.99 $\mu$m H$_2^{18}$O feature.}
\tablenotetext{b}{Peak line flux value of best-fit slab water model divided by the continuum value at that same wavelength, from \citet{Arulanantham25}.}
\tablenotetext{c}{Hot, warm, and cold subscripts refer to best-fit parameters of a 3-component water vapor slab model fit.}
\end{deluxetable*}
\end{rotatetable*}

\begin{figure*}[ht!]
\centering
\includegraphics[width=165mm]{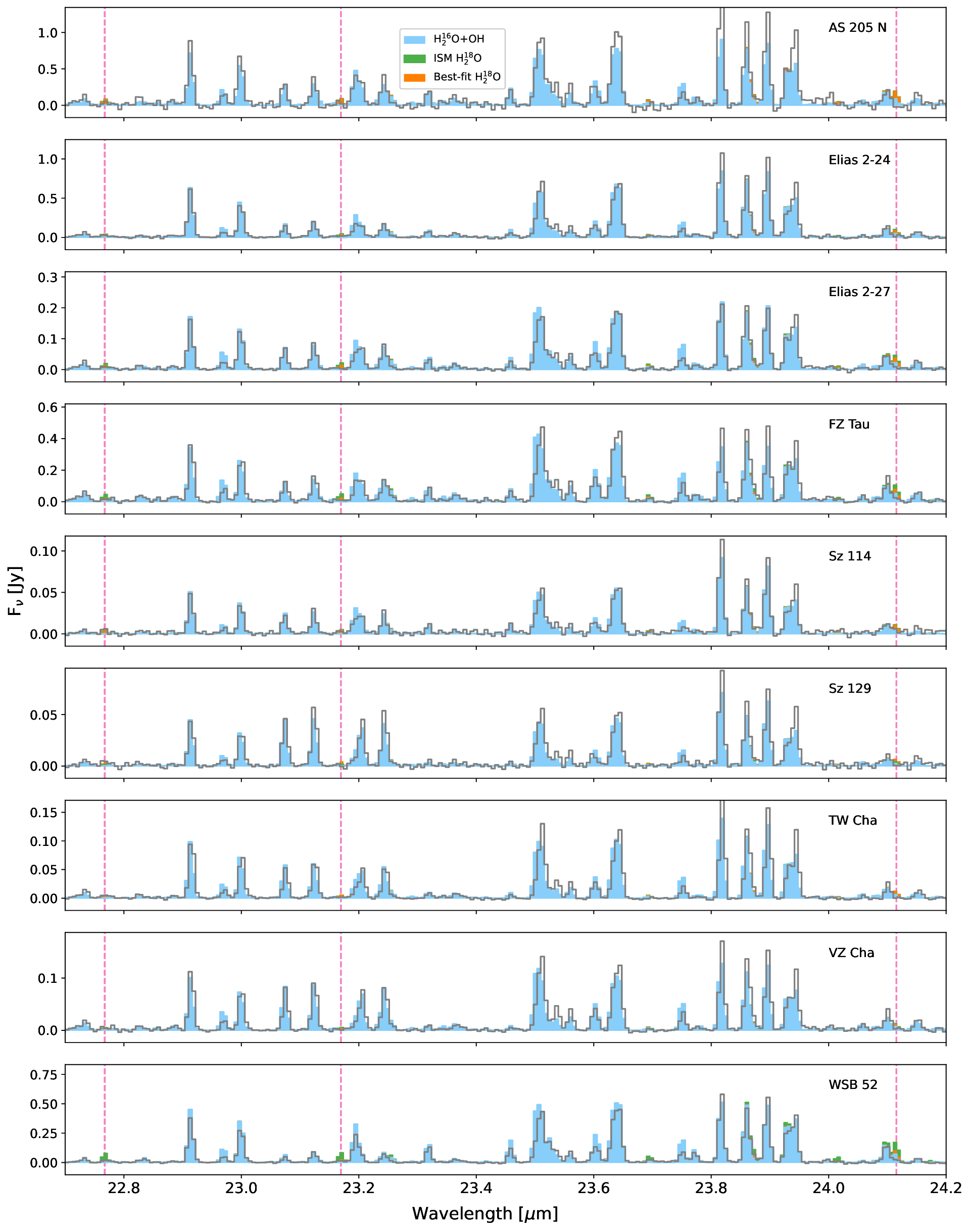}
\caption{Spectra of targets meeting our selection criteria for further analysis (P/C>0.2 and H$_2^{18}$O SNR>2), between 22.7 and 24.2 $\mu$m, along with best-fit 3-component water models (blue; see Table \ref{table:disks}), 3-component H$_2^{18}$O models (green) assuming an ISM $^{16}$O/$^{18}$O ratio of 557, and 3-component H$_2^{18}$O models with a best-fit  $^{16}$O/$^{18}$O ratio (orange).  Vertical dashed lines show H$_2^{18}$O lines from Table \ref{table:stronglines}.
\label{fig:allspectra_short}
}
\end{figure*}

\begin{figure*}[ht!]
\centering
\includegraphics[width=165mm]{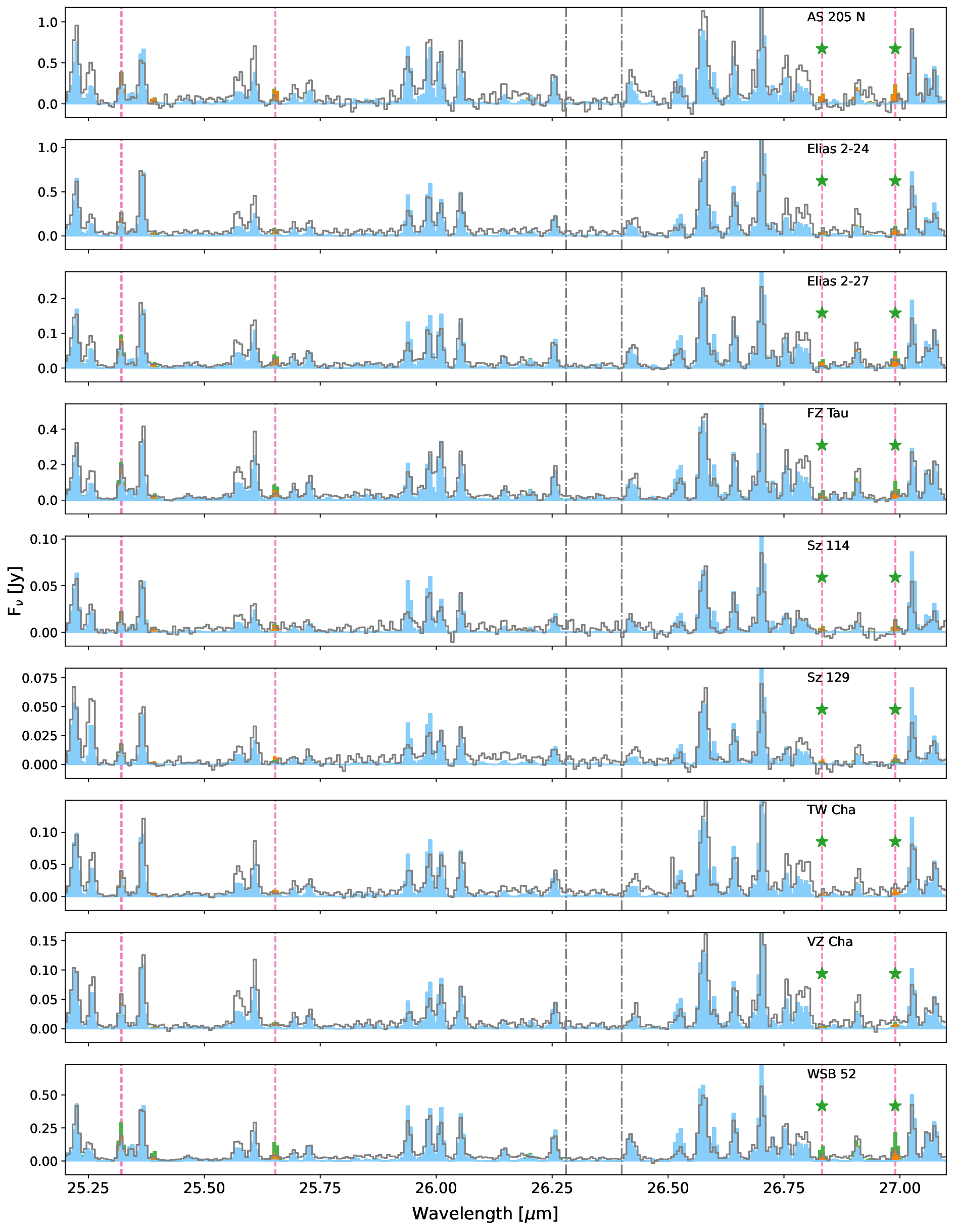}
\caption{Same as Figure \ref{fig:allspectra_short} but for the wavelength range of 25.2 to 27.1 $\mu$m. Gray vertical dash-dot lines mark the location used for determination of the noise level.  Green stars mark the H$_2^{18}$O lines used to fit the H$_2^{16}$O/H$_2^{18}$O ratio. A zoomed-in portion of these features is provided in Figure \ref{fig:allspectra_long_zoom}.
\label{fig:allspectra_long}
}
\end{figure*}

\begin{deluxetable}{c c c c c}
\tablehead{
\colhead{Transition} & \colhead{Wavelength [$\mu$m]} & \colhead{E$_\mathrm{up}/\mathrm{k}$ [K]} & \colhead{A [s$^{-1}$]}}
\startdata
8$_{(7,2)}$ - 7$_{(6,1)}$  &  26.990  &  2265.6  &  20.8 \\
8$_{(7,1)}$ - 7$_{(6,2)}$  &  26.990  &  2265.6  &  20.8 \\
\hline
9$_{(6,3)}$ - 8$_{(5,4)}$  &  26.832  &  2329.5  &  14.9 \\
\hline
8$_{(8,0)}$ - 7$_{(7,1)}$  &  25.653  &  2544.9  &  29.1 \\
8$_{(8,1)}$ - 7$_{(7,0)}$  &  25.653  &  2544.9  &  29.1 \\
\hline
9$_{(7,3)}$ - 8$_{(6,2)}$  &  25.322  &  2581.7  &  22.0 \\
9$_{(7,2)}$ - 8$_{(6,3)}$  &  25.320  &  2581.7  &  22.0 \\
\hline
9$_{(8,2)}$ - 8$_{(7,1)}$  &  24.115  &  2862.2  &  30.5 \\
9$_{(8,1)}$ - 8$_{(7,2)}$  &  24.115  &  2862.2  &  30.5 \\
\hline
9$_{(9,0)}$ - 8$_{(8,1)}$  &  23.169  &  3165.9  &  40.5 \\
9$_{(9,1)}$ - 8$_{(8,0)}$  &  23.169  &  3165.9  &  40.5 \\
\hline
10$_{(8,3)}$ - 9$_{(7,2)}$  &  22.768  &  3213.7  &  32.1 \\
10$_{(8,2)}$ - 9$_{(7,3)}$  &  22.767  &  3213.7  &  32.1 \\
 \enddata
\caption{HITRAN data for strong H$_2^{18}$O lines identified in our modeling. \label{table:stronglines}}
\end{deluxetable}

The three-component H$_2^{18}$O models are shown overlaid on the observed spectra in Figures \ref{fig:allspectra_short}, \ref{fig:allspectra_long}, \ref{fig:nondetections_short}, \ref{fig:nondetections_long}, and \ref{fig:allspectra_long_zoom} in green.  To determine if H$_2^{18}$O is detected in a given spectrum, we estimate the SNR by first measuring the noise level in the relatively water-free region between 26.26 and 26.4 $\mu$m  (see vertical dot-dash lines in Figures \ref{fig:allspectra_long} and \ref{fig:nondetections_long}).  Then, we perform a Gaussian fit to the observed spectrum at the location of the strongest line pair located at 26.99 $\mu$m.  We define the H$_2^{18}$O SNR as the height of the Gaussian divided by the noise level, and show the resultant values in Table \ref{table:disks}.  We note that this relatively strong 26.99 $\mu$m feature is also well-isolated from any strong H$_2^{16}$O emission lines; therefore, although there remain H$_2^{16}$O modeling residuals, these should not affect the emission at this wavelength.

Visual inspection shows that some sources can have marginal detections according to this SNR criterion even though water emission is not particularly strong in the target -- see, for example, DoAr 25 in Figure \ref{fig:nondetections_long}, which has a likely spurious spike near the 26.99 $\mu$m H$_2^{18}$O feature.  Therefore, we use two criteria to define a sample for further analysis of H$_2^{18}$O --- a water peak/continuum ratio (P/C in Table \ref{table:disks}) > 0.2 and H$_2^{18}$O SNR > 2.  Note that we set the criterion at SNR>2, rather than the traditionally utilized SNR>3, so that we can look further into upper limits on H$_2^{18}$O in Section \ref{sec:ratios}.  We do not claim that these are formal detections of H$_2^{18}$O.

\begin{figure*}[h]
\centering
\includegraphics[width=160mm]{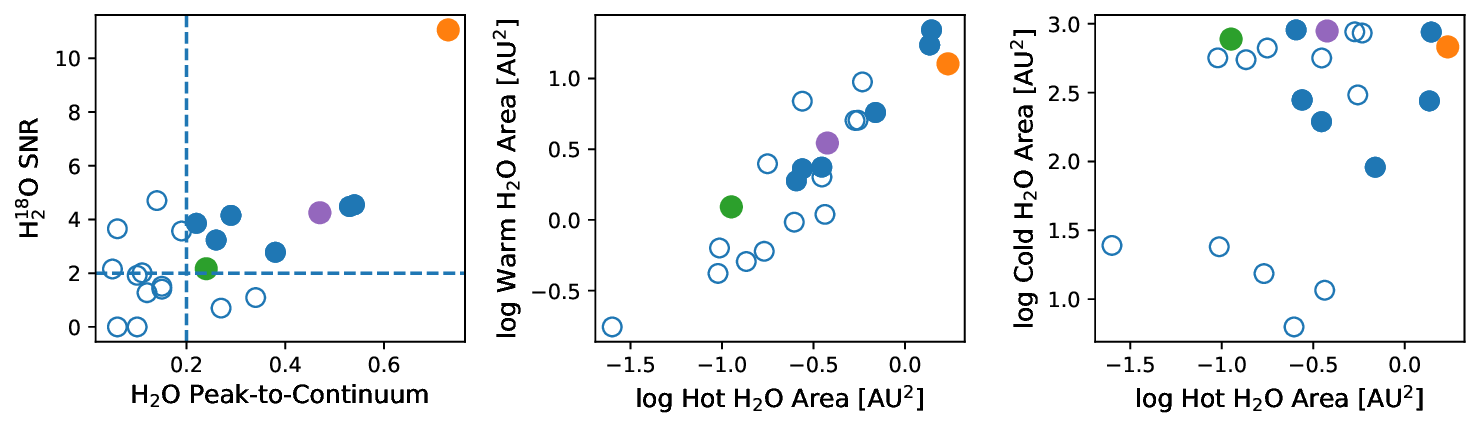}
\caption{Left: H$_2^{18}$O signal-to-noise ratio vs. H$_2$O peak-to-continuum ratio from \citet{Arulanantham25}. Dashed lines mark the cutoffs of 2 and 0.2 for these two parameters, respectively, to select for further analysis.  Here and in the other two panels, filled circles mark sources that meet the cutoff criteria, WSB 52 is highlighted in orange, Sz 129 in green, and TW Cha in purple.  Center: Emitting areas of warm vs. hot H$_2^{16}$O models fit to spectra --- see Table \ref{table:disks}.  Right: Emitting areas of cold vs. hot H$_2^{16}$O models. \label{fig:demographics}}
\end{figure*}

Figure \ref{fig:demographics} shows the criteria for the full water-rich sample, and highlights the nine selected targets meeting the two criteria for more detailed analysis.  The right two panels in Figure \ref{fig:demographics} show the properties of the water models for the full sample and the more selective sample meeting the peak/continuum and H$_2^{18}$O  SNR criteria.  We see that sources meeting the criteria tend to have large warm and hot water emitting areas as compared to the rest of the sample, and we also note that these two areas are highly correlated.  The selection criteria do not correlate with the area of the cold component, or with the temperature or column densities of any of the three components (these correlations are not shown).  

Two targets meet the peak/continuum>0.2 selection criterion for water but only have 2<SNR<3 for the 27 $\mu$m H$_2^{18}$O feature ---  Sz 129 (SNR=2.17) and VZ Cha (SNR=2.78).   Seven targets meet the peak/continuum>0.2 selection criterion for water, and also have SNR>3 on the 27 $\mu$m H$_2^{18}$O feature --- AS 205 N (SNR=3.86), Elias 2-24 (SNR=4.15), Elias 2-27 (SNR=4.55), FZ Tau (SNR=4.48), Sz 114 (SNR=3.24), TW Cha (SNR=4.25) and WSB 52 (SNR=11.06).  Thus, we report marginal detections (SNR$\approx3-5$) of H$_2^{18}$O in 6 disks, and a strong detection of H$_2^{18}$O in one disk: WSB 52.  We provide measured H$_2^{18}$O line fluxes, or upper limits, for the 26.832 and 26.99 $\mu$m lines for these nine targets in Table \ref{table:linefluxes}.  In the next section, we consider whether the spectra allow us to place any constraints on the H$_2^{16}$O/H$_2^{18}$O ratio.

\subsection{Measuring the Isotopologue Ratio}
\label{sec:ratios}
For sources {\it not} meeting our selection criteria for further study, visual inspection of the H$_2^{18}$O models presuming an ISM H$_2^{16}$O/H$_2^{18}$O ratio (see Figures \ref{fig:nondetections_short} and \ref{fig:nondetections_long}, and especially the strongest H$_2^{18}$O feature near 27 $\mu$m) suggests that predicted H$_2^{18}$O emission is similar to, or below, the noise level. Therefore, these spectra are consistent with the water reservoir in the emitting region having an H$_2^{16}$O/H$_2^{18}$O ratio equal to the ISM value, but they do not otherwise strongly constrain this ratio.

For sources {\it meeting} our selection criteria for further study, visual inspection of the green (ISM ratio) models in Figures \ref{fig:allspectra_long} and  \ref{fig:allspectra_long_zoom} suggests that some H$_2^{18}$O models seem consistent with the data, while others appear to over-predict the observed H$_2^{18}$O emission --- see especially FZ Tau and WSB 52.  Therefore, there is not clear evidence for enhancement of H$_2^{18}$O and, in some cases, this modeling framework instead implies that the H$_2^{16}$O/H$_2^{18}$O ratio must be {\it greater} than the ISM value (i.e., the observed water is actually depleted in H$_2^{18}$O).

To investigate this observation quantitatively, we create a grid of models for each target in which the H$_2^{18}$O features are assumed to emit from the same three reservoirs as in the prior modeling, but the H$_2^{16}$O/H$_2^{18}$O ratio is allowed to vary.  We do not try to independently fit the slab model parameters for H$_2^{18}$O due to the small number of (marginally)-detected emission lines of H$_2^{18}$O.  H$_2^{18}$O temperatures and emitting areas are taken to be the same as in Table \ref{table:disks}, but the H$_2^{18}$O column density is modified according to the H$_2^{16}$O/H$_2^{18}$O ratio. We focus our model fitting on the two strongest H$_2^{18}$O features at 26.832 and 26.99 $\mu$m.  Additionally, to counteract issues with continuum determination (for example, see how the derived continuum can dip below 0 for AS 205 N near the 26.832 $\mu$m feature in Figures \ref{fig:allspectra_long} and \ref{fig:allspectra_long_zoom}), we calculate integrated fluxes for the 26.832 and 26.99 $\mu$m features and compare model fluxes to these calculated fluxes to determine model residuals (and $\chi^2$ values).  Modeling the integrated line fluxes is also an appropriate approach since these emission lines are spectrally unresolved \citep{Banzatti25}.  We then use the $\chi^2$ goodness-of-fit parameter to determine the best-fit H$_2^{16}$O/H$_2^{18}$O ratio.
 
\begin{figure}[h]
\centering
\includegraphics[width=75mm]{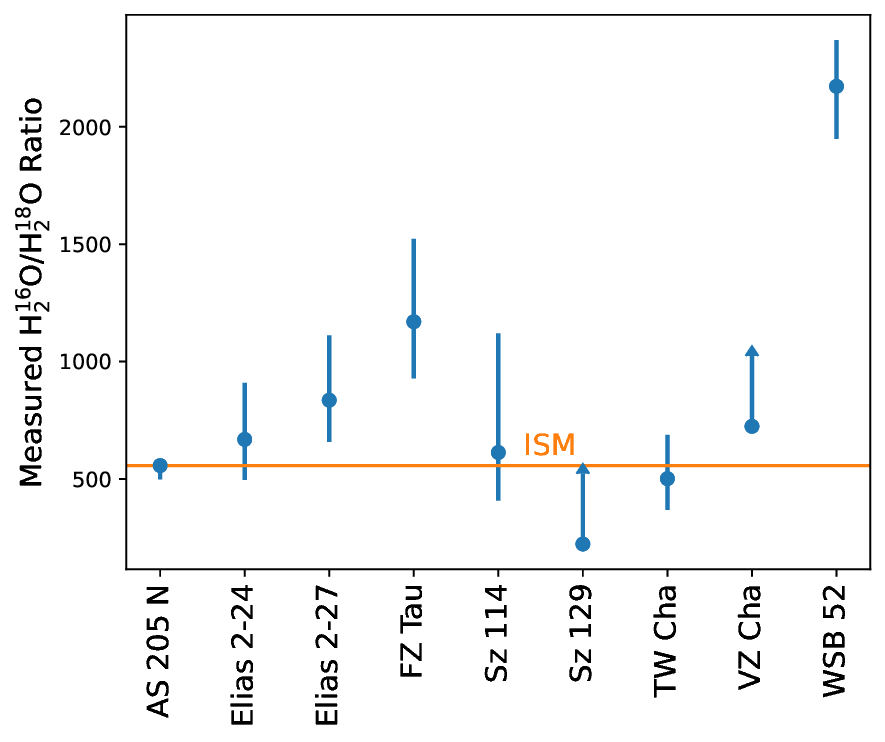}
\caption{Best-fit H$_2^{16}$O/H$_2^{18}$O ratios, or lower limits.  A horizontal line shows the ISM value of 557.  
\label{fig:ratios}}
\end{figure}

 Best-fit  H$_2^{16}$O/H$_2^{18}$O ratios are provided in Table \ref{table:ratios} and Figure \ref{fig:ratios}, and best-fit models are shown in orange in Figures \ref{fig:allspectra_short}, \ref{fig:allspectra_long} and \ref{fig:allspectra_long_zoom}.  When the H$_2^{18}$O SNR is $<3$, we report the ratio as a lower limit.

\begin{deluxetable}{l c}
\tablecaption{Best-fit H$_2^{16}$O/H$_2^{18}$O ratios for sources meeting selection criteria described in Section \ref{sec:detections}\label{table:ratios}}
\tablehead{\colhead{Source}&\colhead{H$_2^{16}$O/H$_2^{18}$O\tablenotemark{a}}}
\startdata
AS 205 N  &  557 $^{+ 23 }_{- 59 }$ \\
Elias 2-24  &  668 $^{+ 241 }_{- 173 }$  \\ 
Elias 2-27  &  836 $^{+ 276 }_{- 177 }$ \\ 
FZ Tau  &  1170 $^{+ 353 }_{- 242 }$  \\  
Sz 114  &  613 $^{+ 507 }_{- 205 }$ \\
Sz 129  &  >223 \\ 
TW Cha  &  501 $^{+ 187 }_{- 132 }$ \\
VZ Cha  &  >724 \\ 
WSB 52  &  2172 $^{+ 197 }_{- 224 }$ \\ 
\enddata
\tablenotetext{a}{Best-fit isotopologue ratio and 1$\sigma$ error bars, found by minimizing the $\chi^2$ goodness-of-fit parameter.  Lower limits are provided when the SNR of the H$_2^{18}$O feature is <3.}
\end{deluxetable}

\section{Discussion}
\subsection{Are the H$_2^{16}$O/H$_2^{18}$O ratios truly anomalous?}
\label{sec:anomalous}
Only one disk with detected H$_2^{18}$O --- TW Cha --- and one disk with SNR$\sim$2 --- Sz 129 --- have best-fit H$_2^{16}$O/H$_2^{18}$O ratios lower than the ISM value, implying possible enhancement in H$_2^{18}$O. (We discuss these sources further in Section \ref{sec:sz129}.)  Other best-fit H$_2^{16}$O/H$_2^{18}$O ratios (Table \ref{table:ratios}) are either consistent with or higher than the standard ISM value of 557 \citep{Milam05}.  We consider here whether this could be an artifact of our modeling process, or an excitation effect, rather than a true measurement of H$_2^{16}$O/H$_2^{18}$O in the disk atmospheres.

\begin{figure*}[h]
\centering
\includegraphics[width=160mm]{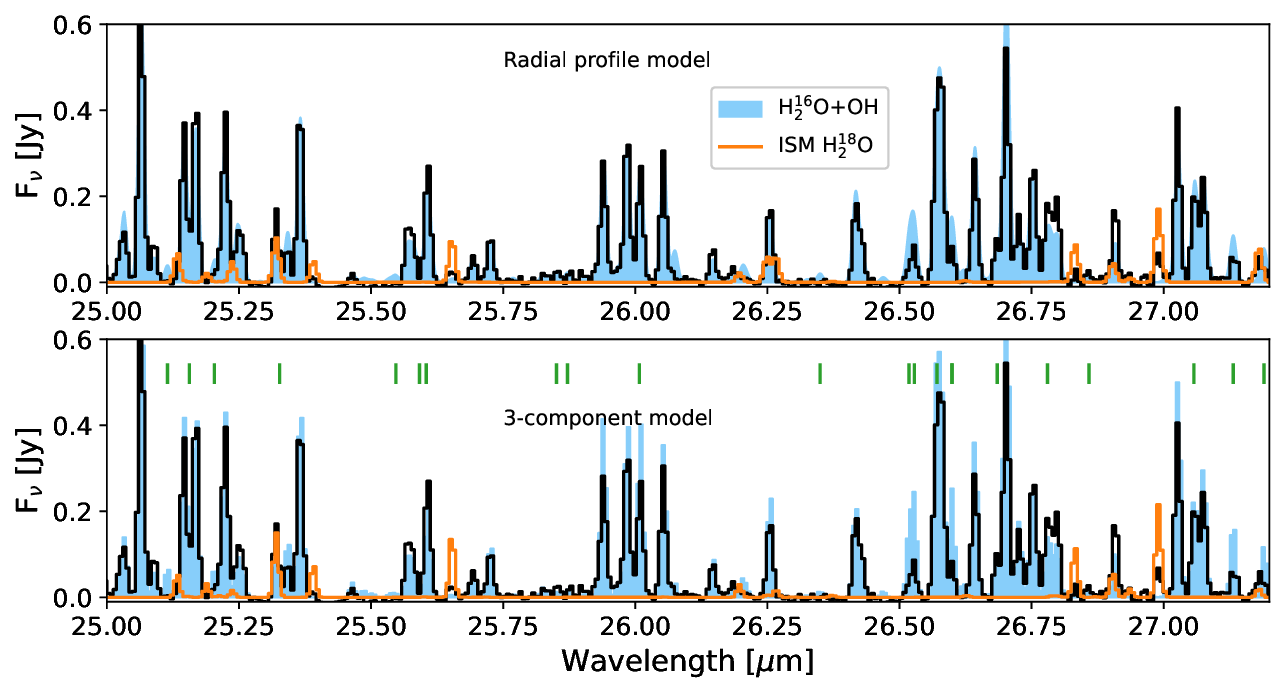}
\caption{Comparison of WSB 52 water model fits using a radial profile model (top) and a 3-component slab model (bottom).  The observed spectrum is in black, the modeled H$_2^{16}$O+OH spectrum is in blue, and the modeled H$_2^{18}$O spectrum (which assumes an ISM isotopologue ratio) is in orange.  Green vertical lines mark locations of strong vibrationally excited transitions.
\label{fig:twomodel_comparison}}
\end{figure*}

For our nominal models, we used a 3-component slab model to fit the H$_2^{16}$O spectrum.  Therefore, we first investigate whether using a model with radial temperature and column density gradients might produce different best-fit H$_2^{16}$O/H$_2^{18}$O ratios.  In Figure \ref{fig:twomodel_comparison}, we compare our initial model fits for WSB 52 --- the disk with the strongest H$_2^{18}$O detection --- with those utilizing a radial profile model.  We utilize the approach described in \citet{Romero-Mirza24}, in which the disk is modeled as consisting of a series of rings extending from 0.1 to 10 AU with log radial spacing, and H$_2^{16}$O temperature and column density are described by power laws $T = T_0 \left( \frac{r}{0.5 \mathrm{au}}\right)^{-\alpha}$ and $N = N_0 \left( \frac{r}{0.5 \mathrm{au}}\right)^{-\beta}$, respectively.  We compare modeled line fluxes to measured line fluxes for 51 pure rotational H$_2^{16}$O lines --- a subset of the 81 isolated H$_2^{16}$O lines identified in \citet{Banzatti25} (see their Tables 5 and 6) with high-quality fits.

Running a grid of models and minimizing $\chi^2$, we find a best-fit model with $N_0 = 1.3\times 10^{18}\, \mathrm{cm}^{-2}$, $T_0 = 794$ K, $\alpha=0.7$ and $\beta=-1.9$.  We also assume a disk inclination of $i=54^\circ$ \citep{Huang18} and a distance of 135 pc \citep{Gaia23}.  A portion of the resultant model spectrum is shown in Figure \ref{fig:twomodel_comparison}.  We then also produce an H$_2^{18}$O model with the same parameters, but $N_0$ reduced by a factor of 557 (the ISM H$_2^{16}$O/H$_2^{18}$O ratio).  The orange curves in Figure \ref{fig:twomodel_comparison} show that this ISM ratio model similarly over-predicts the H$_2^{18}$O emission compared to observations.  Therefore, changing to a radially-varying water model does not apparently change the conclusion that some H$_2^{16}$O/H$_2^{18}$O ratios are larger than the ISM value.

We also consider whether vertical temperature gradients, unaccounted for in our nominal slab models, might produce an anomalous H$_2^{16}$O/H$_2^{18}$O ratio.  A proposed argument might be as follows: the H$_2^{16}$O is more optically thick than the H$_2^{18}$O and, therefore, the H$_2^{16}$O emission probes higher and hotter regions of the externally-heated disk atmosphere as compared to H$_2^{18}$O. Utilizing the H$_2^{16}$O model parameters might then tend to over-predict the true H$_2^{18}$O temperature and, therefore, its emitted flux.  In fact, Figure \ref{fig:allspectra_short} does show that the nominal H$_2^{18}$O models tend to overpredict the line fluxes for the shorter-wavelength (higher energy) lines, suggesting that the H$_2^{18}$O may be better characterized by a lower temperature.

The logic of the proposed argument in its simplest form may not be self-consistent, however.  By making the assumption that H$_2^{16}$O and H$_2^{18}$O arise from the same reservoir, we are implicitly asking what the H$_2^{16}$O/H$_2^{18}$O ratio is in the observed H$_2^{16}$O reservoir.  And in that observable reservoir, H$_2^{18}$O appears to be under-abundant. Therefore, having the H$_2^{18}$O also emit from an additional, cooler, reservoir below this H$_2^{16}$O reservoir does not negate the need for a high H$_2^{16}$O/H$_2^{18}$O in the region above, as that reservoir is already over-predicting observed fluxes. 

Nevertheless, we look into this further by asking --- by what $\Delta $T would we need to alter our H$_2^{18}$O model in order to match observed line fluxes, and is this $\Delta $T consistent with expected disk atmospheric gradients?  We find that if the warm and hot H$_2^{18}$O reservoir temperatures (in the 3-component slab model) are reduced by 150 K, the modeled H$_2^{18}$O fluxes match observed fluxes.  However, by utilizing the disk models produced by \citet{Calahan22}, we see that the emitting region for the 26.99 $\mu$m H$_2^{18}$O emission lines, which are moderately optically thick, overlaps with the layer at which H$_2^{16}$O self-shielding becomes effective.  Therefore, the emitting regions for the observed H$_2^{18}$O and the bulk of the observable H$_2^{16}$O are similar, and unlikely to have large differences in temperature.

\begin{figure*}[h]
\centering
\includegraphics[width=160mm]{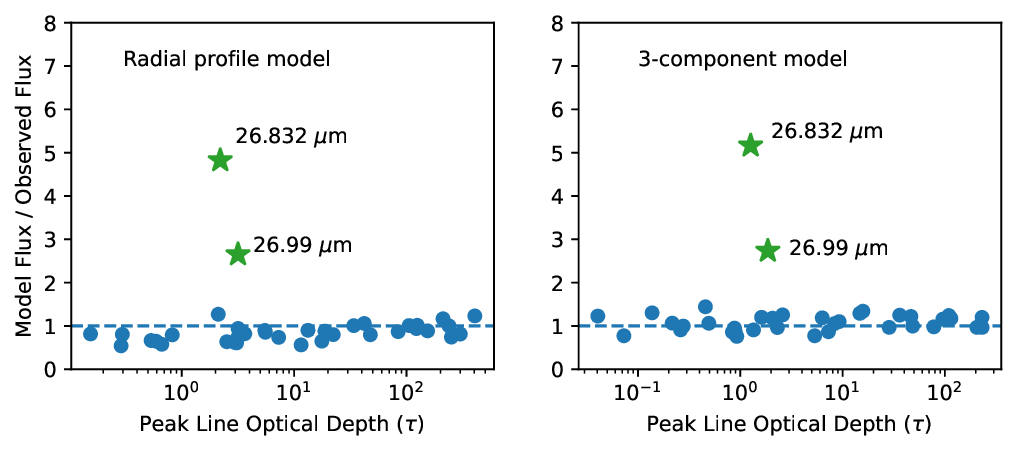}
\caption{Ratio of modeled to observed line fluxes for the radial profile model (left) and three-component slab model (right).  Blue circles show H$_2^{16}$O lines while green stars show  H$_2^{18}$O lines.  Labels show wavelengths of the two H$_2^{18}$O features.  The H$_2^{16}$O lines selected for this plot have E$_\mathrm{up}/$k$<5000$ K, and are pure rotational transitions.  The H$_2^{18}$O models assume an ISM isotopologue ratio. 
\label{fig:flux_comparison}}
\end{figure*}

We can also perform a simple test of this idea by asking whether H$_2^{16}$O emission lines with lower optical depths also tend to be over-predicted.  In Figure \ref{fig:flux_comparison}, we plot the ratio of model-predicted to observed line fluxes for the two fit H$_2^{18}$O features (at 26.832 and 26.99 $\mu$m) and all pure rotational H$_2^{16}$O lines with E$_\mathrm{up}$/k$<5000$ K.  We find that modeled H$_2^{16}$O lines for a wide range of optical depths are consistent with observed lines to within a factor of two, and do not show the discrepancy observed for the H$_2^{18}$O features. \footnote{Note that visual comparison between the H$_2^{16}$O models and data might suggest larger model vs. data discrepancy --- for example, see the discrepancy in the H$_2^{16}$O line complex near 26.52 $\mu$m in Figure \ref{fig:allspectra_long}.  However, we find that these larger discrepancies occur for vibrationally excited lines, and not for pure rotational lines.  }

However, these arguments ignore non-thermal excitation effects. As discussed in \citet{Banzatti25},  the excitation of water vapor in disk atmospheres likely involves a combination of collisional and radiative excitation. Thanks in part to the high optical depth of many of the ground state pure rotational transitions, the critical density, including photon escape probabilities, is lower for the pure rotational lines in the ground state as compared to the bending mode ``hot band'' pure rotational lines also traced by  MIRI-MRS at wavelengths beyond 17 $\mu$m. The rovibrational bending manifold emission near 6 $\mu$m has higher critical densities still. For the H$_2^{18}$O isotopologue emission, the lower peak optical depths of the lines will result in higher critical densities than those for the main isotopologue, which will result in less efficient excitation of the H$_2^{18}$O states observed here. The quantitative differences in the rotational excitation of H$_2^{16}$O and H$_2^{18}$O in disk atmospheres will require detailed non-LTE simulations, which are beyond the scope of this work.


\subsection{Implications of measured H\textsubscript{2}\textsuperscript{16}O/H\textsubscript{2}\textsuperscript{18}O ratios}
In this section, we discuss some of the implications of our H$_2^{18}$O observations.  

\subsubsection{Confirmation of measured water column densities}
Although H$_2^{16}$O emission from protoplanetary disk atmospheres has been modeled now for more than a decade and has consistently found water vapor column densities of $\sim10^{18}$ cm$^{-2}$ for typical full disks around T Tauri stars \citep[e.g.][]{Carr08,Carr11, Salyk11,Banzatti23a,Temmink24}, many observed H$_2^{16}$O lines are optically thick. Therefore, one might wonder if the  H$_2^{16}$O column density could be underestimated.  Comparison of primary and secondary isotopologues offer an alternative way to estimate gas column densities, as the ratio of optically thick (primary isotopologue) to optically thin (secondary isotopologue) lines decreases as optical depths rise \citep[e.g.][]{Grant23,Salyk25}.  If H$_2^{16}$O emission modeling had been underestimating water column densities, we would expect to find {\it stronger} than predicted H$_2^{18}$O line emission.  Instead, the many non-detections and marginal detections of H$_2^{18}$O show no evidence for underestimation of H$_2^{16}$O column densities.

\subsubsection{Sz 129 and TW Cha}
\label{sec:sz129}
 Sz 129 and TW Cha are consistent with best-fit H$_2^{16}$O/H$_2^{18}$O values lower than the ISM value (albeit consistent with the ISM value to within error bars/limits).  This could imply enhancement in H$_2^{18}$O relative to the ISM. As highlighted in Figure \ref{fig:demographics}, among the sources in our subsample selected for further study, Sz 129 has the smallest hot and warm H$_2^{16}$O emitting areas, but a relatively large cold H$_2^{16}$O emitting area.  In addition, \citet{Arulanantham25} note that Sz 129 is an outlier with relatively low HCN line flux, but relatively high cold water content.  Interestingly, this disk has a 10 AU mm-wave cavity \citep{Huang18}.  The water reservoir probed in this disk may therefore have different properties than that probed in the other disks in our subsample.

TW Cha shows no unusual water emission properties, sitting in the middle of the population shown in Figure \ref{fig:demographics}.  It does, however, have a $\sim$30 AU millimeter-wave cavity \citep{Arulanantham25}. 

\subsubsection{WSB 52}
Our highest SNR detection of the 26.99 $\mu$m H$_2^{18}$O emission feature occurs for the disk around WSB 52.  As highlighted in Figure \ref{fig:demographics}, this disk has the largest hot H$_2^{16}$O emitting area and the largest H$_2^{16}$O peak-to-continuum ratio.  Given that the H$_2^{18}$O emission is dominated by warm and hot components (see Figure \ref{fig:contribution_plot}), the large line-to-continuum ratio of the H$_2^{16}$O might simply predict a large line-to-continuum ratio for H$_2^{18}$O as well, making it more readily detectable.  

There is no obvious reason why WSB 52 might have a particularly high  H$_2^{16}$O emitting area.  WSB 52's disk and stellar properties do not seem anomalous in any particular way, including having an estimated water ``pebble flux'' that is unremarkable in comparison to other disks observed with MIRI-MRS \citep{Krijt25}.  It is, however, one of the smaller mm-wave disks in the JDISCS sample (radius of 32 AU; \citealp{Huang18,Arulanantham25}).  It also displays ``wide-angle conical outflow structures'' consistent with a disk wind (Narang et al., 2025, in preparation).

\subsubsection{Scenarios that fail to explain enhanced H$_2^{16}$O/H$_2^{18}$O ratios}
If the high H$_2^{16}$O/H$_2^{18}$O ratio in the water-emitting region observed for some disks in our sample is not caused by modeling effects (including needing to account for non-LTE excitation), then our modeled H$_2^{16}$O/H$_2^{18}$O ratios would imply true higher-than-ISM values  in that region.  In this section, we begin by considering a few possible explanations that we ultimately do {\it not} find plausible to explain the measured ratios.

 Firstly, we note that there is no obvious way to produce H$_2^{16}$O/H$_2^{18}$O ratios above the ISM value via optical depth effects.  In order for isotope-selective photodissociation to enhance  H$_2^{18}$O in the disk atmosphere, the CO must self-shield at a layer above where the H$_2$O shields.  This allows the CO to ``donate'' its preferential abundance of free $^{18}$O to the production of water vapor in the dominant water-emitting region.  Therefore, one might posit that, in disk regions with enhanced H$_2$O/CO ratios, the H$_2$O self-shielding region could lie above the CO self-shielding region.  However, unlike for CO, which is dissociated in distinct narrow bands, for water vapor, H$_2^{16}$O serves as a broad UV shield \citep[e.g.][]{Fillion01,Bethell09,Adamkovics14} which would shield the H$_2^{18}$O as effectively as the H$_2^{16}$O. Thus, the self-shielding of water vapor is not expected to change the local H$_2^{16}$O/H$_2^{18}$O ratio; furthermore, it would shield all isotopologues of CO below that level, erasing any possibility of isotope-selective dissociation of the CO.

Similarly, the models of \citet{Calahan22} demonstrate that, in order for H$_2^{18}$O enhancements to be {\it observed}, the H$_2^{18}$O line emitting region must overlap with the region of enhanced $^{18}$O, Therefore, one might posit that changes in disk structure could reduce the extent of overlap, or eliminate it altogether. However, if these regions do not overlap, the observed water emission would simply reflect the un-modified (ISM) value.



Another possibility is that the intrinsic $^{16}$O/$^{18}$O ratios in some of these disks are simply different from the ISM ratio.  However, measurements of $^{16}$O/$^{18}$O throughout the Galaxy seem to peak at $\sim$600 \citep{Wilson94}, while we find ratios as high as 2200 (see Table \ref{table:ratios}).  In addition, the typical spread in this ratio at a given Galactocentric radius is no more than $\sim$100 \citep{Wilson94}.  Therefore, observations of observed Galactic $^{16}$O/$^{18}$O ratios do not support this scenario.  However, it should be noted that these Galactic values all come from surveys of H$_2$CO \citep{Wilson94,Gardner81,Schuller85},  and, therefore, make the assumption that the isotope ratios in this molecule reflect the bulk ISM ratios.

\subsubsection{Potential scenarios to explain enhanced H$_2^{16}$O/H$_2^{18}$O ratios}


We consider here two possible processes that might enhance the gas phase H$_2^{16}$O/H$_2^{18}$O in the disk itself.  

In the first scenario, water is initially enhanced in H$_2^{18}$O in the gas phase, according to the CO selective-photodissociation models of the pre-stellar core \citep{Lee08} or the outer disk \citep{Lyons05}, but the now H$_2^{18}$O-enriched water is subsequently removed from the observable emitting region of the disk.  One possible removal mechanism would be incorporation of the H$_2^{18}$O-enriched water into planetesimals.  Another possible removal mechanism could be drift of the H$_2^{18}$O-rich water onto the parent star.  In both cases, what remains in the gas phase is CO relatively depleted in $^{18}$O.  This ``light'' CO, if broken apart, perhaps by He+ ions generated via X-ray ionization \citep{Bergin14,Furuya14}, could serve as a source for the gas-phase formation of $^{18}$O-depleted water vapor.

In a second scenario, pebble drift near the midplane may fractionate the ice and gas in a mass-dependent fashion. At temperatures near 185 K, for example, \citet{Smith03} find that the sublimation rate of H$_2^{18}$O from crystalline water ice is $\sim$9\% slower than for the parent isotopologue (the D/H ratio in the vapor is depleted by factors closer to 50\% in these kinetic experiments). Further, in experiments on silicate dust plus ice mixtures, \citet{Moores12} find that the vapor above such mixtures is depleted in HDO by factors of $\sim$two at similar temperatures, suggesting that the fractionation of H$_2^{18}$O observed by \citet{Smith03}, near temperatures of 185 K, might become significant under the much colder protostellar disk conditions near the water snowline.


From laboratory work on realistic ice mixture compositions \citep[e.g.][]{Kipfer24}, sublimation from dust/ice aggregates will begin at temperatures near 140--150 K and continue through higher temperatures as pebbles drift radially inward. By combining the laboratory experiments and a simple two-step model with kinetic fractionation factors for sublimation followed by equilibrium condensation at vapor/ice ratios of $\sim$1/5 (values suggested by models of the two-dimensional vapor transport near snowlines, \citealp{Wang25}), the resulting water vapor is predicted to be depleted by a factor of two in $^{18}$O --- i.e., within the range of isotopic ratios observed in our MRS data. While simplistic, such an approximate model does suggest that large effects are possible near the snowlines of protoplanetary disks.

Further evidence of significant water vapor/ice fractionation is provided by submillimeter observations of the outbursting source V883 Ori, in which water vapor generated by the sublimation of ice beyond the low-accretion rate snowline is extremely enriched in Deuterium \citep{Tobin23}.  Evidence of a ``retreating snowline'' in V883 Ori \citep{Wang25} is also consistent with our proposed cycle of sublimation followed by re-condensation. More detailed modeling that incorporates isotopic-selective dissociation, vertical vapor/ice transport and radial pebble drift, and gas-grain interactions and cycling could test the various ice/vapor scenarios outlined here. 



Note that a convenient feature of these scenarios is that the first is mass-independent, while the second is mass-dependent. Therefore, as in meteoritic studies, observations of H$_2^{17}$O (in addition to H$_2^{18}$O) in protoplanetary disks could be the key to unlock the physics behind the observed water isotopologue ratios.   In addition, both scenarios predict evolutionary changes to the H$_2^{16}$O/H$_2^{18}$O ratio, with slight differences between them.  In the first, H$_2^{18}$O is enhanced early on via isotope-selective photodissociation, but later depleted as the first-generation H$_2$O is removed and subsequently replaced by the $^{18}$O-poor second-generation H$_2$O.  In the second scenario,  H$_2^{18}$O is depleted with time as cycles of sublimation and condensation occur at the snowline, and this depletion should be linked to pebble evolution.    These evolutionary scenarios could potentially be tested by measuring ratios in disks of different ages, and with different degrees of pebble evolution, although development of detailed models is also needed.

\section{Conclusions}
We searched for H$_2^{18}$O in 22 JDISCS targets, focusing in particular on a subset of nine with high water line/continuum ratios and possible evidence for H$_2^{18}$O emission lines.  We report marginal detections (SNR=3--5) in six of these disks, and a more significant detection (SNR=11) in the disk around WSB 52.   

We modeled the H$_2^{16}$O emission from these nine disks using 3-component slab models, then used ISM H$_2^{16}$O/H$_2^{18}$O ratios to compare predicted and observed H$_2^{18}$O fluxes, or upper limits.  Although the strongest H$_2^{18}$O lines lie at the long-wavelength end of the MRS range, their fluxes are dominated by warmer components of water emission models, and not by the cold component that potentially traces solid water transport \citep{Banzatti23b}.

We fit the strongest H$_2^{18}$O emission features with a model allowing a varying H$_2^{16}$O/H$_2^{18}$O ratio.  We find that two disks are consistent with H$_2^{16}$O/H$_2^{18}$O<ISM in the water-emitting region (although still consistent with ISM values within error bars/limits).   The remaining seven are consistent with  H$_2^{16}$O/H$_2^{18}$O$\geq$ISM in that region (three of the seven consistent with ISM to within 1 $\sigma$).  This result appears robust to changes in many modeling assumptions, although further investigation into non-LTE excitation is warranted.   H$_2^{16}$O/H$_2^{18}$O ratios $>$ISM values are inconsistent with models of isotope-selective photodissociation in the inner few AU of a protoplanetary disk atmosphere\citep{Calahan22}.  As alternatives, we propose either that the $^{18}$O-rich water could be removed from the emitting region and replaced with  $^{18}$O-poor water via gas-phase reactions powered by chemical processing of CO, or that the $^{18}$O-poor water we see could have been affected by mass-dependent fractionation via a cycle of snowline sublimation/condensation.

To our knowledge, the $\sim$11$\sigma$ detection of H$_2^{18}$O in WSB 52 is the first such detection reported in a planet-forming region to date.  This result highlights the unprecedented sensitivity provided by JWST and MIRI-MRS.  However, if H$_2^{18}$O had been present in the disk atmosphere at ISM abundances relative to H$_2^{16}$O, the H$_2^{18}$O emission should have been detectable in a larger number of disks; thus, the detection of oxygen isotopologues in planet-forming regions is more challenging than expected, even for JWST.  Nevertheless, even the upper limits, and small number of detections reported here place meaningful constraints on models of isotope fractionation in planet-forming regions beyond our solar system.  One avenue for future study is the use of spectrographs (like the proposed PRIMA FIRESS instrument; \citealp{Pontoppidan25}) that study the far-IR, where our models predict stronger H$_2^{18}$O features, and more sensitivity to colder, potentially transport-dominated, reservoirs of water vapor.

\begin{acknowledgments}
The authors wish to especially thank Iouli Gordon and Robert Hargreaves for helpful conversations about the HITRAN and HITEMP databases.  This project was supported by the STScI grant JWST-GO-01584.001-A ``A DSHARP-MIRI Treasury survey of Chemistry in Planet-forming Regions''.  A portion of this research was carried out at the Jet Propulsion Laboratory, California Institute of Technology, under a contract with the National Aeronautics and Space Administration (80NM0018D0004). This research used the SpExoDisks Database at spexodisks.com. 
\end{acknowledgments}
\clearpage

\appendix

\section{HITEMP/HITRAN comparison}
\label{sec:hitemp_hitran_comparison}
\begin{figure*}[ht!]
\centering
\includegraphics[width=140mm]{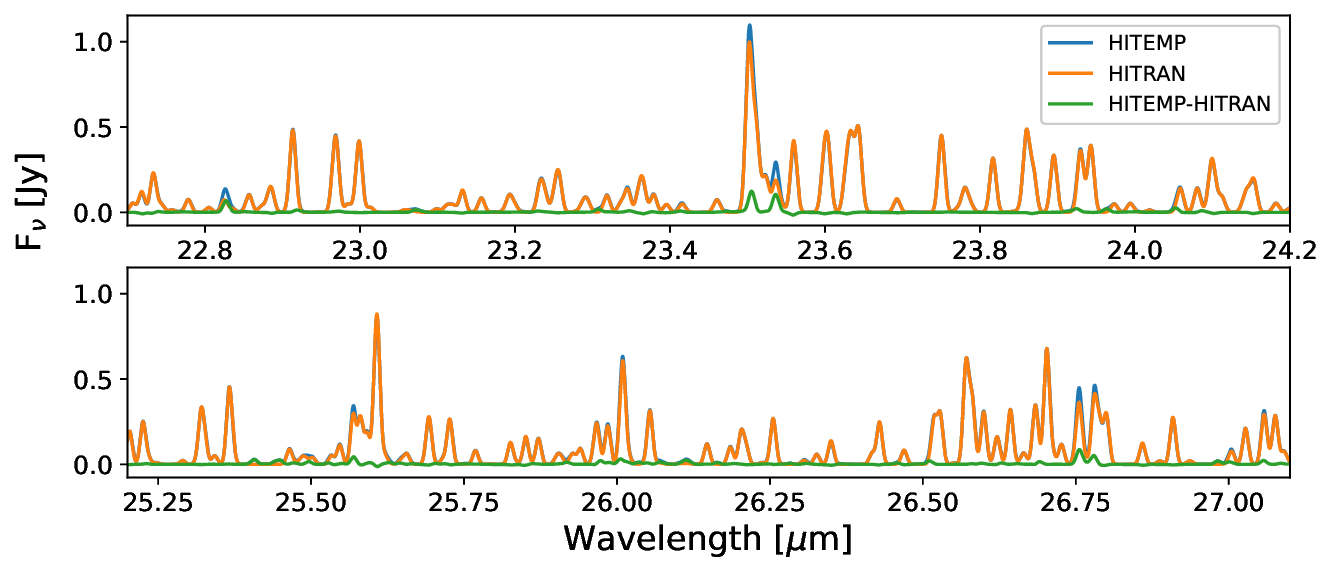}
\caption{ A comparison of slab models produced using our restricted HITEMP line list, and the HITRAN database. Slab models have temperatures of 1000 K, column densities of 10$^{18}$ cm$^{-2}$ and emitting areas of $\pi$ (1 AU)$^2$, at a distance of 150 pc.  \label{fig:hitemp_hitran_comparison} }
\end{figure*}

Figure \ref{fig:hitemp_hitran_comparison} shows a comparison of slab models produced using our restricted HITEMP water list (see Section \ref{sec:h216o_models}) and the HITRAN database.  We show the regions from 22.7-24.2 $\mu$m and 25.2-27.1 $\mu$m, which are analyzed in the remainder of this work.  Discrepancies at 23.505 and 23.535 are due to a known bug in the creation of the HITRAN 2020 linelist, which sometimes omits strong lines; this bug is fixed for the HITRAN 2024 release (Louli Gordon, priv comm).  Other discrepancies are caused by the normal restrictions imposed on the HITRAN linelist \citep{Gordon22}.

\section{Spectra not selected for further analysis, and zoomed-in spectra}
\label{sec:nondetections}
\begin{figure*}[ht!]
\centering
\includegraphics[width=140mm]{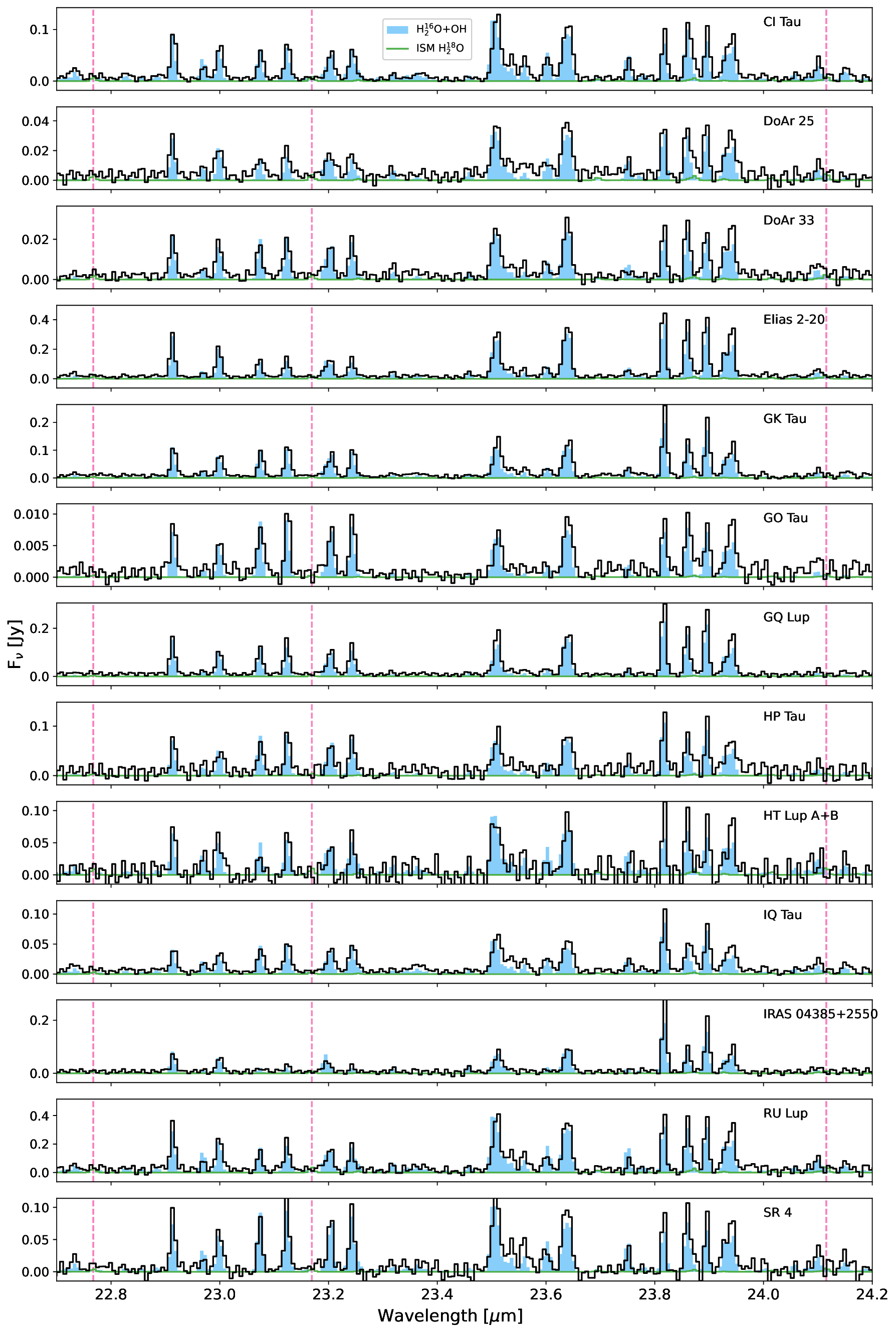}
\caption{Spectra of targets not meeting our selection criteria, between 22.7 and 24.2 $\mu$m, along with best-fit 3-component water models (blue; see Table \ref{table:disks}) and 3-component H$_2^{18}$O models (green) assuming an ISM $^{16}$O/$^{18}$O ratio of 557.  Pink vertical dashed lines show H$_2^{18}$O lines from Table \ref{table:stronglines}.
\label{fig:nondetections_short}
}
\end{figure*}

\begin{figure*}[ht!]
\centering
\includegraphics[width=140mm]{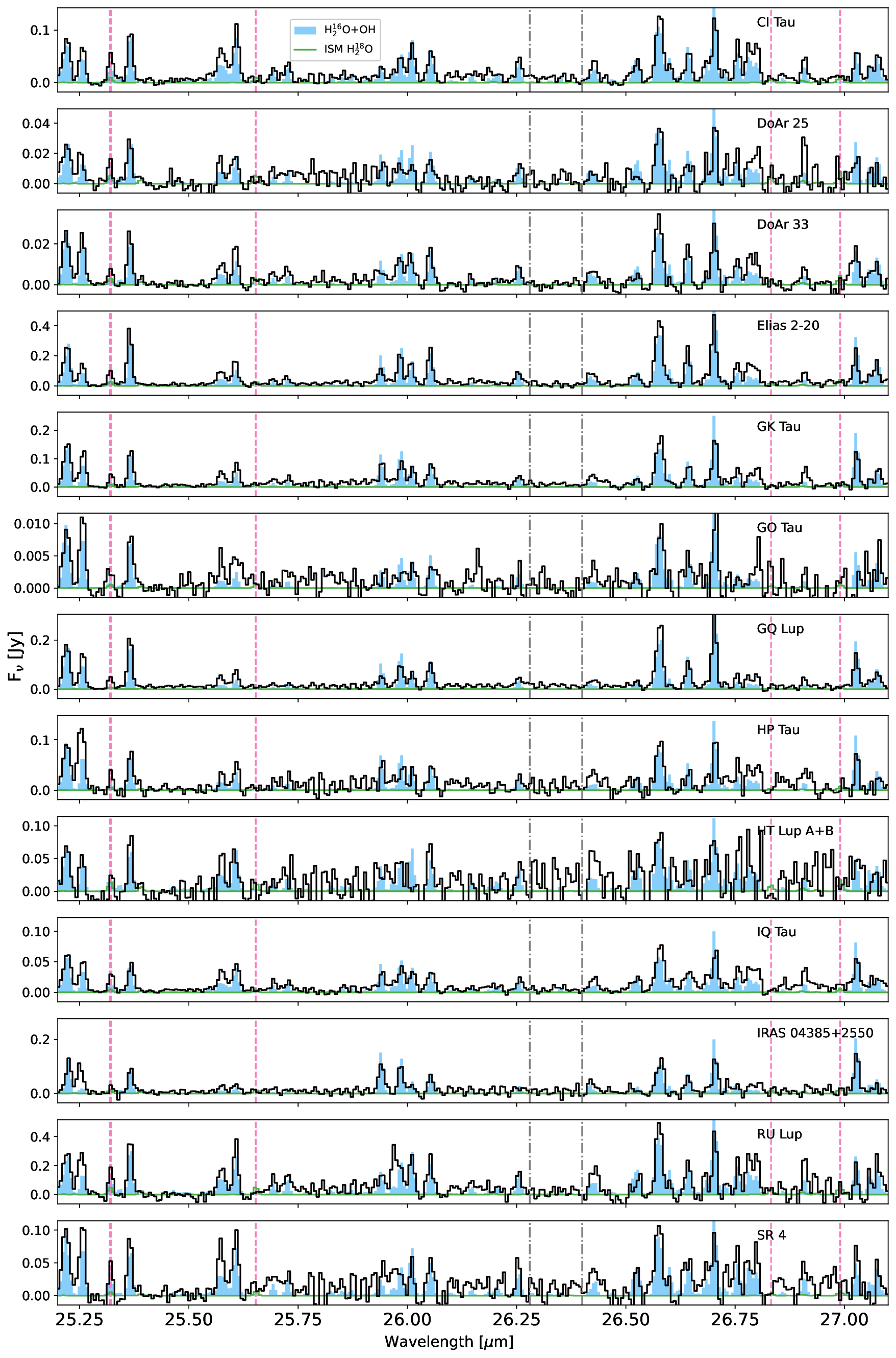}
\caption{Same as Figure \ref{fig:nondetections_short} but for the wavelength range of 25.2 to 27.1 $\mu$m. 
\label{fig:nondetections_long}
}
\end{figure*}

Figures \ref{fig:nondetections_short} and \ref{fig:nondetections_long} show observed spectra in the H$_2^{18}$O-emitting regions for sources not selected for further analysis.  

\begin{figure}[ht!]
\centering
\includegraphics[width=75mm]{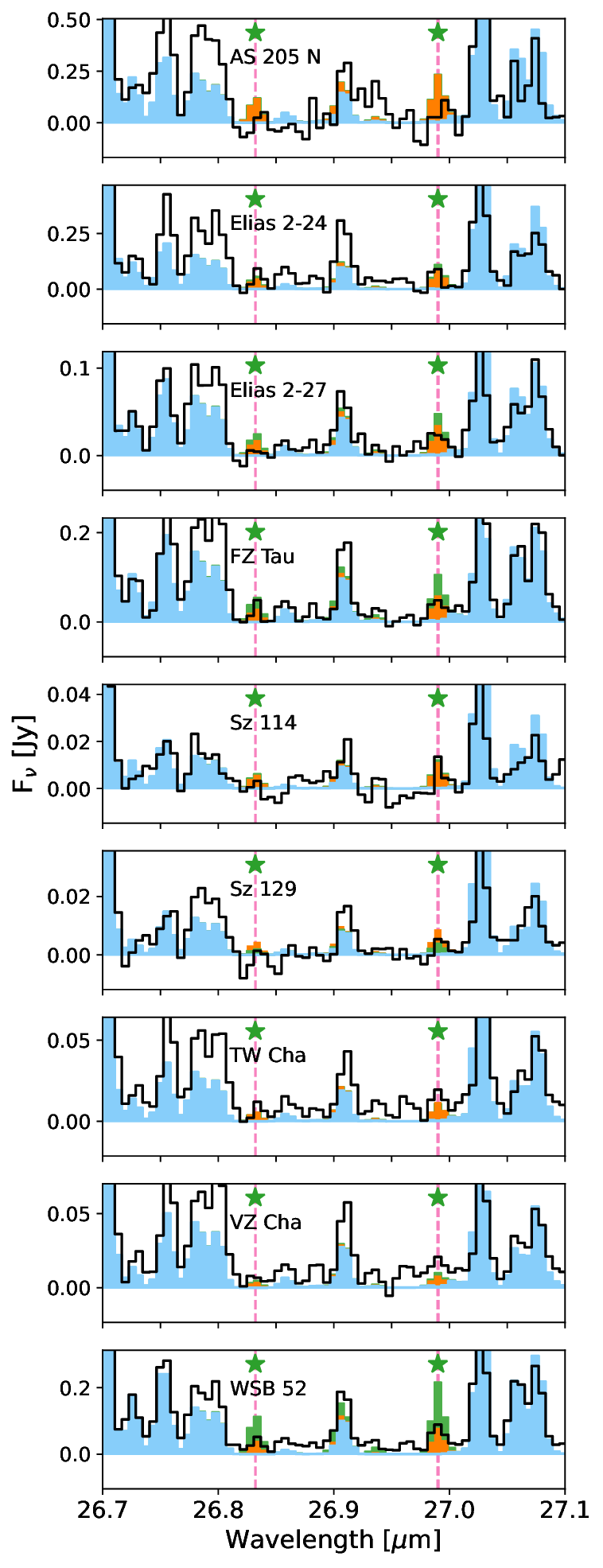}
\caption{Same as Figure \ref{fig:allspectra_long}, but zoomed in on the H$_2^{18}$O emitting region near 27 $\mu$m.  Best-fit 3-component water models are shown in blue, 
H$_2^{18}$O models assuming an ISM $^{16}$O/$^{18}$O ratio of 557 in green, and  H$_2^{18}$O models with a best-fit  $^{16}$O/$^{18}$O ratio in orange. Green stars mark the two lines which fluxes are used to find the best-fit H$_2^{16}$O/H$_2^{18}$O ratio.
\label{fig:allspectra_long_zoom}
}
\end{figure}

 Figure \ref{fig:allspectra_long_zoom} shows a zoomed-in plot of the H$_2^{18}$O-emitting region near 27 $\mu$m for the sources selected for further study.

\section{Measured line fluxes}
\label{sec:linefluxes}
Table \ref{table:linefluxes} provides the measured line fluxes, or 3$\sigma$ upper limits, for the H$_2^{18}$O emission features at 26.832 and 26.99 $\mu$m, respectively, for our subsample selected for further study.

\begin{deluxetable*}{l c c}
\tablecaption{Measured line fluxes for the 
H$_2^{18}$O features at 26.832 and 26.99 $\mu$m. \label{table:linefluxes}}
\tablehead{\colhead{Source}&\colhead{Flux}&\colhead{Flux}\\
&\colhead{[$10^{-19}$ W m$^{-2}$]} & \colhead{[$10^{-19}$ W m$^{-2}$]}\\
&\colhead{26.832 $\mu$m} & \colhead{26.99 $\mu$m}}
\startdata
AS 205 N  &  46.8 $\pm$ 21.4  &  165.0 $\pm$ 42.8 \\
Elias 2-24  &  31.9 $\pm$ 11.1  &  46.0 $\pm$ 11.1 \\
Elias 2-27  &  3.4 $\pm$ 1.9  &  17.4 $\pm$ 3.8 \\
FZ Tau  &  15.1 $\pm$ 3.5  &  31.6 $\pm$ 7.1 \\
Sz 114  &  2.5 $\pm$ 1.8  &  6.0 $\pm$ 1.8 \\
Sz 129  &  $<$ 8.7  &  $<$ 11.6 \\
TW Cha  &  3.3 $\pm$ 1.3  &  6.2 $\pm$ 1.5 \\
VZ Cha  &  $<$ 1.3  &  $<$ 15.0 \\
WSB 52  &  11.8 $\pm$ 2.6  &  40.4 $\pm$ 3.6 \\
\enddata
\end{deluxetable*}

\end{document}